\documentclass[10pt,prl,twocolumn,natbib,showkeys] {revtex4}

\usepackage[dvips]{graphicx}

\begin{document}

\title{Fine tuning of phase qubit parameters for optimization of fast single-pulse readout}

\author{\firstname{Leonid S.} \surname{Revin}}
\author{\firstname{Andrey L.} \surname{Pankratov}}
\email{alp@ipm.sci-nnov.ru}

\affiliation{Institute for Physics of Microstructures of RAS,
GSP-105, Nizhny Novgorod, 603950, Russia}

\begin{abstract}

We analyze a two-level quantum system, describing the phase qubit, during a single-pulse readout process by a numerical solution of the time-dependent Schroedinger equation. It has been demonstrated that the readout error has a minimum for certain values of the system`s basic parameters. In particular, the optimization of the qubit capacitance and the readout pulse shape leads to significant reduction of the readout error. It is shown that in an ideal case the fidelity can be increased to almost 97$\%$ for 2 ns pulse duration and to 96$\%$ for 1 ns pulse duration.
\end{abstract}
\date{\today}
\maketitle
\newpage

\begin{figure}[h]
\resizebox{1\columnwidth}{!}{
\includegraphics{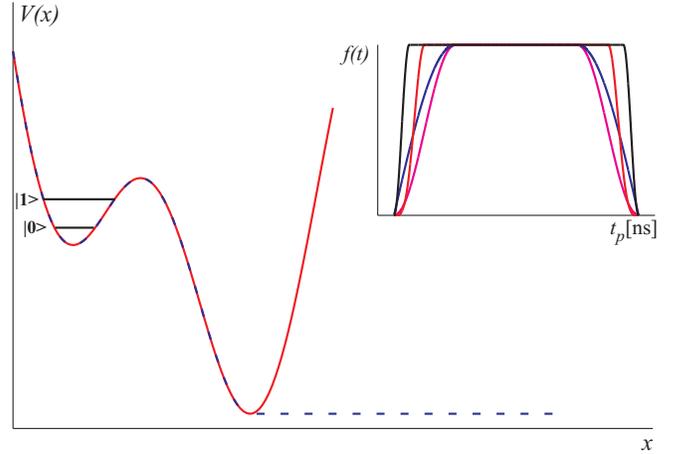}}
{\caption{The profile of a bistable potential. Solid curve - the original potential, dashed curve - the potential with enlarged deep well to simulate the effect of damping. Inset: the considered trapezoid-like pulses with walls, from top to bottom, $\sin^2(8\pi t/t_p)$, $\sin^2(4\pi t/t_p)$, $\sin(2\pi t/t_p)$, $\sin^2(2\pi t/t_p)$.} \label{fig1}}
\end{figure}

In recent years superconducting Josephson junction circuits have attracted a considerable interest as promising devices for quantum computations \cite{q1}-\cite{q4}. To speed-up the readout of a qubit state and improve its fidelity, the single-pulse readout technique has been suggested \cite{five}-\cite{three}. According to this technique, a measurement of a flux-biased phase qubit, described by a shallow potential well with two energy levels $\left|0\right>$ and $\left|1\right>$ (see Fig. \ref{fig1}), is performed by subjecting the qubit to a pulse driving. It leads to lowering the barrier between the qubit "left" and "right" potential wells, so the system will tunnel from state $\left|1\right>$ with a probability close to one, while state $\left|0\right>$ remains intact. However, for small readout times the achieved fidelity was rather low, of order 70-80 $\%$, which was explained by different sources of decoherence. Recently, it has been demonstrated that the coherent Rabi oscillations remain nearly unaffected by thermal fluctuations up to temperatures of 1K \cite{temp} (i.e., until the energy of thermal fluctuations $kT$ becomes comparable with the energy level spacing $\hbar \omega$ of the qubit), so without degrading the already achieved coherence times, phase qubits can be operated at temperatures much higher than those reported so far. This may signal that relatively large readout errors of practical devices \cite{one} can be attributed to non-optimal readout of the qubits rather than quantum and thermal fluctuations. In \cite{six} it has been demonstrated that the readout error has a minimum as function of both amplitude and duration of the readout pulse, as well as the qubit capacitance. On one hand it allows to obtain high fidelity for fixed pulse duration by designing the qubit with a proper capacitance, and by changing the pulse amplitude. On the other hand it is possible to decrease the readout time, making it much smaller than the qubit coherence time and suppressing the effect of thermal fluctuations \cite{seven}. As it has been understood \cite{two},\cite{six}, the pulse shape has a strong influence on different error probabilities. Also, it is believed that the main source of error during the qubit readout is due to incomplete discrimination between the two quantum states, which is determined by the depth of a shallow potential well. However, the investigation versus external magnetic field (which is changing the potential well depth and can be varied during an experiment), and also existence of the optimal readout pulse shape were not studied. The main goal of the paper is to find the optimal values of qubit parameters for very fast readout of order 1 ns.

We describe a phase qubit by the following potential (shown in Fig. 1, solid curve) \cite{six}:
\begin{equation}
V(x,t)=E_J\left\{(x-\varphi(t))^2/{2\ell}-\cos x\right\}. \label{V}
\end{equation}
 Here $E_J=I_C\hbar/2e$ is the Josephson energy, $x$ is the Josephson phase, $e$ is the electron charge, and $\hbar$ is the Planck constant. The qubit parameters are taken the same as in \cite{one2}-\cite{three}: the critical current $I_C=1.7\rm {\mu A}$, the inductance of the ring $L=0.72$ nH and the capacitance $C=700$ fF, correspond to $\ell=2eI_C L/\hbar=3.71$, ${2e^2}/{\hbar C}=0.6933\times 10^9$ Hz, $E_J/\hbar=I_C/2e=5.31\times 10^{12}$ Hz. It is convenient to introduce the "inverse capacitance" $D={2e^2}/{\hbar C}\times 10^{-9}$ Hz, and express the time in nanoseconds. The dimensionless external magnetic flux $\varphi(t)=2\pi[a_0+Af(t)]$ consists of two components: the dc component $a_0$, which determines the depth of the shallow well, and the driving readout pulse with an amplitude $A$ and pulse shape $f(t)$, the trapezoid-like function, which grows and drops by different laws (see the inset of Fig. \ref{fig1}). We note that $t_p$ is defined as the full width of the pulse at zero level, rather than the full width at half maximum.

 The number of the discrete energy levels can be characterized by the crude estimated value:
  \begin{equation}
    N_l={\triangle U_l}/{\hbar\omega_l}, \label{N}
    \end{equation}
    where $\triangle U_l$	is the depth of the left well (the energy difference between the potential maximum and minimum) and $\omega_l$ is the classical oscillation frequency near the left-well bottom (the "plasma frequency"): $\omega_l=\sqrt{E_J(1/\ell+\cos x_l)/m}$, where $x_l$ corresponds to the left-well bottom and $m=\hbar/(2D)$ is the effective mass.

  Let us consider the readout error $N$, which is the sum of two probabilities, $P_{10}$ not to tunnel during the pulse action from the state $\left|1\right>$, and $P_{01}$ to tunnel from the state $\left|0\right>$ (i.e. $N=P_{10}+P_{01}$, while the fidelity $F=1-N$). The investigation is performed via computer simulation of the Schroedinger equation and is focused on the readout error $N$ versus the pulse amplitude and the shape, as well as the depth of a shallow potential well.

The Schroedinger equation for the wave function $\Psi(x,t)$ has the following form:
\begin{equation}
i \frac{\partial\Psi(x,t)}
{\partial t}=-\frac{2e^2}{\hbar C}\frac{\partial^2\Psi(x,t)}{\partial x^2}+\frac{V(x,t)}{\hbar}\Psi(x,t). \label{Sch}
\end{equation}
The Eq. (\ref{Sch}) does not take into account either damping or noise, and in the frame of Eq. (\ref{Sch}) we will only study the error occuring due to incomplete discrimination of the quantum states $\left|0\right>$ and $\left|1\right>$. As it has been done in \cite{six}, to prevent the repopulation error, we consider a modified potential, which differs from Eq. (\ref{V}) that at the bottom of the deep well the potential does not grow up and remains constant to the right really far away, see Fig. 1, dashed curve. Therefore in the boundary conditions $\Psi(c,t)=\Psi(d,t)=0$ we take $c=-3$, $d=797$, while the right-well minimum is located at $x_r\approx 6$ and $x_l\approx 1.4$. We have checked that further increase of this range does not change the results.

For fast single-pulse readout system we take the pulse duration $t_p = 2$ ns. A special feature of the quantum system is the impossibility to use rectangular readout pulses \cite{six}, which in the classical case leads to minimal noise-induced errors \cite{seven}. In the qubit a rectangular pulse leads to nonadiabaticity of the tunneling event, which results in lifting to higher eigenstates and considerable increase of the probability to tunnel from $\left |0\right>$ state and thus to much larger values of $N$ than for all other pulse shapes. As it will be shown, to obtain the minimal error we should use a compromise shape, close to a meander, but not yet leading to the discussed effect.

\begin{figure}[h]
\resizebox{1\columnwidth}{!}{
\includegraphics{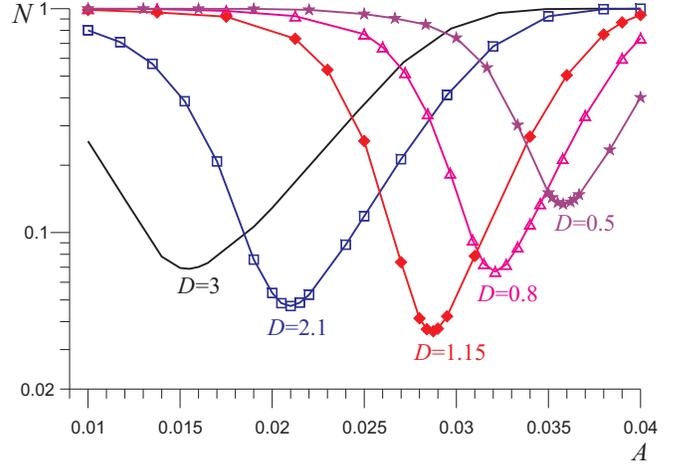}}
{\caption{The readout error $N$ versus the pulse amplitude $A$ for the pulse with walls $\sin^2(4\pi t/t_p)$ and $a_0=0.81$.} \label{fig2}}
\end{figure}
First, let us take a sine-trapezoid function which grows and drops for $t\le t_p/8$ and $t\ge 7t_p/8$ by $\sin^2(4\pi t/t_p)$. The readout error $N$ versus the pulse amplitude $A$ is presented in Fig. 2 for different values of the inverse capacitance $D$ and the shift of a potential barrier $a_0=0.81$. It is seen that $N$ has a minimum depending on the amplitude, as well as depending on the parameter $D$. 
Decrease of $D\to 0$ leads to increase in the number of energy levels and to decrease in the distance $\hbar\omega_l\to 0$ between them (see Eq. (\ref{N}), where $\omega_l \sim \sqrt{D}$), which complicates the discrimination between the states $\left|0\right>$ and $\left|1\right>$. If $\hbar\omega_l$ is large, both ground and excited states become too close to the barrier top, so the tunneling from both states may occur even without the driving pulse, leading to large readout error. Therefore, since both limits of large and small $D$ lead to large readout errors, there must be some optimal value of the inverse capacitance, leading to the minimal readout error $N(A,D)$.

Note that to find the minimal $N(A)$ for a fixed value of $D$ it is not necessary to plot the whole curve $N(A)$. Changing the inverse capacitance leads to a shift of the curves on a specific number $A_s$, e.g. increase of $D$ by 0.3 leads to a decrease $N_{min}(A)$ on $A$ to 0.0037. Thus, one can predict the location of $N(A)$ minimum for different $D$, that speeds-up either calculations or measurements.

\begin{figure}[h]
\resizebox{1\columnwidth}{!}{
\includegraphics{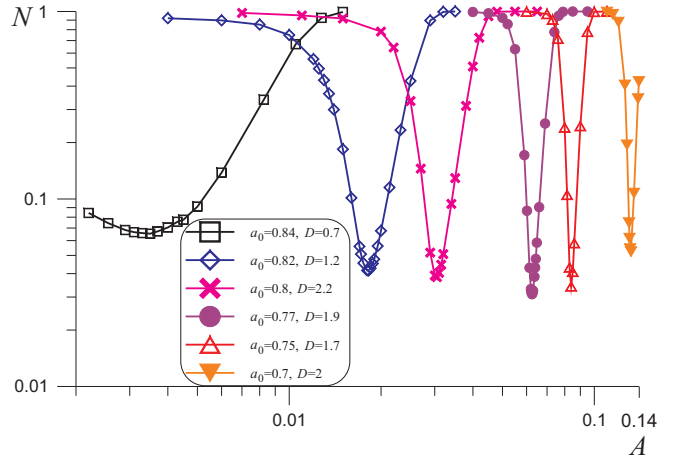}}
{\caption{The readout error $N$ versus the pulse amplitude $A$ for the pulse with walls $\sin^2(4\pi t/t_p)$ with different inverse capacitances $D$ and dc magnetic field $a_0$.} \label{fig3}}
\end{figure}
The barrier height $\triangle U_l$ depends on the external flux $\varphi(t)$, particularly it is determined by the dc component $a_0$. If the well is too shallow, the variation of the parameter $D$ does not seriously change the small number of the discrete levels $N_l$ (Eq. (\ref{N})) and the tunneling error is still large. If this well is deep enough, the value $N_l$ remains large and discrimination between two nearby states $\left|0\right>$ and $\left|1\right>$ is complicated. The optimal curves with minimal readout error $N_{min}(A,D)$ for different values of external shift $a_0$ are presented in Fig. 3. It demonstrates the absolute minimum $N_{min}(A,D,a_0)$=0.031 for $a_0$=0.77, $D$=1.9 Hz, $A$=0.0625 (fidelity $F$=0.969). The location of this minimum corresponds to some value between five and six energy levels inside the shallow well, which is larger than for $a_0$=0.81 in \cite{six}. The fast qubit state measurement requires that the maximum external flux $\varphi_1=2\pi[a_0+A]$ changes the potential so that the state $\left|1\right>$ is very close to the barrier top. It corresponds to $N_l$ slightly larger than unity. So we can facilitate the optimization process finding a range of $A$ values for the fixed parameters $a_0$ and $D$. For example, for $a_0$=0.82 and $D$=1.2 Hz the minimal error corresponds to the values of $A$ from 0.0166 (for the condition $N_l\le$ 1.1) to 0.0236 (for the condition $N_l\ge$ 1). While for $a_0$=0.7 and $D$=2 Hz, $N_{min}$ is shifted to $A$ from 0.131 to 0.139. These results are confirmed by numerical solutions shown in Fig. 3. Note that this estimate is valid not only for pulse shapes considered in this paper, but even for the triangular and linear ramp pulses.

The second pulse we consider is the sine-trapezoid function which grows and drops for $t\le t_p/4$ and $t\ge 3t_p/4$ by $\sin^2(2\pi t/t_p)$. Using the developed algorithm, the curves with minimal readout error $N_{min}(A,D)$ were found for different constant magnetic field components $a_0$ (Fig. 4, solid curves). For comparison the optimal curves for the pulse with walls $\sin^2(4\pi t/t_p)$ and the same $a_0$ (with their best inverse capacitance $D$) are also presented by dashed curves. Relative to the previous pulse, in this case the absolute minimum shifts toward smaller values of $a_0$ and larger $A$, while the tunneling error $N$ increases to about 0.034. So, reducing the flat part of the sine-trapezoid function leads to worse results.

\begin{figure}[h]
\resizebox{1\columnwidth}{!}{
\includegraphics{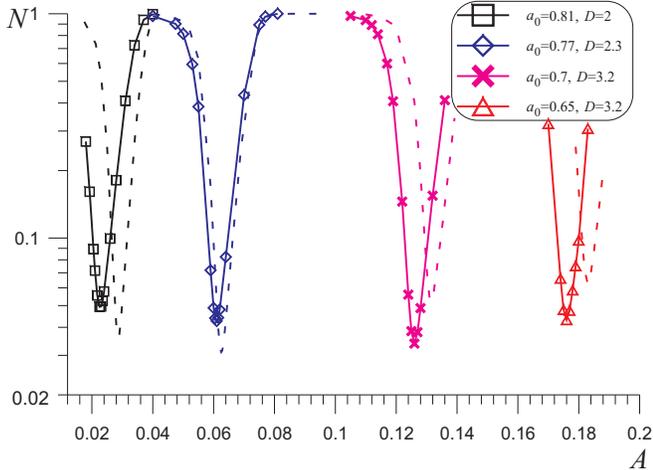}}
{\caption{The readout error $N$ versus the pulse amplitude $A$ for the pulse with walls $\sin^2(2\pi t/t_p)$ with different $D$ and $a_0$ (solid curves). The curves for the pulse with walls $\sin^2(4\pi t/t_p)$ with the same $a_0$ and their own $D$ are given by dashed curves for comparison.} \label{fig4}}
\end{figure}

Let us increase the width of the flat part of the pulse to $14t_p/16$ value. The readout error $N$ versus amplitude $A$ for dc magnetic field $a_0$=0.81 is presented in Fig. 5 (solid curves - pulse with walls $\sin^2(8\pi t/t_p)$, dashed curves correspond to $\sin^2(4\pi t/t_p)$ for the same parameters). As expected, if the pulse shape is more close to a rectangular one, the effect of exitation of $\left|0\right>$ state is more noticeable: there are several local minima of $N$ (for $D$ = 1.1, $N$ = 0.044, and for $D$ = 2.1, $N$ = 0.045). But the absolute value of the error in this case is larger than for the sine-trapezoid pulse with walls $\sin^2(4\pi t/t_p)$.

\begin{figure}[h]
\resizebox{1\columnwidth}{!}{
\includegraphics{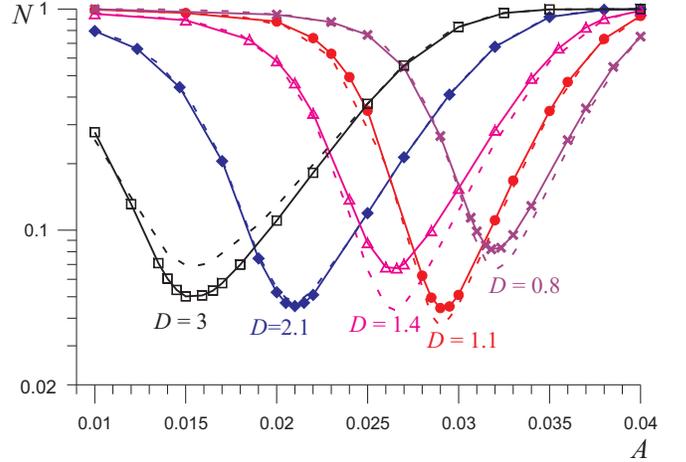}}
{\caption{The readout error $N$ versus the pulse amplitude $A$ for the pulse with walls $\sin^2(8\pi t/t_p)$ and $a_0=0.81$ (solid curves). The dashed curves correspond to the pulse with walls $\sin^2(4\pi t/t_p)$ with the same values of $D$ and $a_0$.} \label{fig5}}
\end{figure}
Thus, we have found that for the sine-trapezoid pulse $\sin^2(4\pi t/t_p)$ with the width $t_p$=2 ns the readout error can be reduced to $N$=0.031 and this value is obtained for the inverse capacitance $D$=1.9 Hz. Therefore, considering the fixed set of experimental parameters of \cite{one2}-\cite{three}, the reduction of $C$ from 700 fF to the range from 450 fF to 150 fF must lead to the highest fidelity. It has been demonstrated \cite{six} that, with decrease in duration, the minimum $N$ is shifted towards larger $A$, but the value of tunneling  error increases significantly. For $t_p$=1 ns, sine-trapezoid function with walls $\sin^2(8\pi t/t_p)$, $a_0$ = 0.77, $A$ = 0.057 and $D$ = 3.4, we achieve $N\approx 0.038$, which leads to the increase of the readout error in about 20 $\%$ only in comparison with 2 ns pulse. Without the pulse shape and dc flux optimization for $t_p$=2 ns $N$ was 0.053 \cite{six}, and the difference of the errors for 1 ns and 2 ns durations was about 40-50 $\%$.

Finally, let us consider a task, where the qubit capacitance $C$ is selected out of the optimal range. In this case one can set some different pulse shape to maximize the fidelity. For example, let us take the inverse capacitance as in \cite{one2}-\cite{three}, $D$=0.6933, corresponding to $C$=700 fF. Here the highest fidelity $F$=94.8 $\%$ is achieved by using smoother pulse shapes, such as sine-trapezoid pulses with walls $\sin^2(2\pi t/t_p)$ and $\sin(2\pi t/t_p)$, while $\sin^2(4\pi t/t_p)$ leads to larger readout error.

In conclusion, we have performed computer simulations of the fast single-pulse readout process of a two-state quantum system. It has been demonstrated that the minimization of the readout error can be achieved by variation of the depth of a shallow potential well of a qubit. Considering the concrete parameters of existing qubit designs \cite{one2} - \cite{three}, it is recommended to decrease the qubit capacitance down to 150-450 fF. Further improvement of the qubit fidelity can be reached during an experiment by adjustment of the dc magnetic field and the readout pulse amplitude. It is demonstrated that there is an optimal pulse shape minimizing the readout error, which for $C=450$ fF is close to sine-trapezoid function with walls $\sin^2(4\pi t/t_p)$. For larger values of the capacitance smoother pulses can lead to the maximal fidelity. Finally, the performed optimizations allowed to reach almost 97$\%$ fidelity for 2 ns pulses and to 96$\%$ fidelity for 1 ns pulses.
While in our model several important sources of decoherence are
not taken into account, the obtained values of the fidelity seem
to be the highest observed for so short pulses (in particular, the readout error was decreased by a factor of 2 in comparison with Ref. \cite{six}), so we believe that the described optimizations can help to improve the fidelity of real devices.

 The work was supported by RFBR (projects 09-02-00491 and 08-02-97033).

\end{document}